\documentclass[aps,pre,onecolumn,groupedaddress]{revtex4}

\usepackage{graphicx} 
\usepackage{dcolumn} 
\usepackage{bm}
\newcommand{\ignore}[1]{} 
\usepackage{epsfig} 
\usepackage{color}
\usepackage{hyperref}
\usepackage{amsmath,amssymb}

\begin{document}

\title{Modeling and analysis of social phenomena: challenges and possible research directions}

\author{Federico Vazquez}
\email{fede.vazmin@gmail.com}
\affiliation{Instituto de C\'alculo, FCEyN, Universidad de Buenos Aires and Conicet, Intendente Guiraldes 2160, Cero + Infinito, Buenos Aires C1428EGA, Argentina.}
\homepage{https://fedevazmin.wordpress.com}

\date{\today}

\begin{abstract}
This opening editorial aims to interest researchers and encourage novel research in the closely related fields of sociophysics and computational social science.  We briefly discuss challenges and possible research directions in the study of social phenomena, with a particular focus on opinion dynamics.  The aim of this special issue is to allow physicists, mathematicians, engineers and social scientists to show their current research interests in social dynamics, as well as to collect recent advances and new techniques in the analysis of social systems.
\end{abstract}

\maketitle

The field of sociophysics, which combines tools and methods from statistical physics to investigate social phenomena, has grown tremendously in the last two decades and is becoming an established research discipline.  Understanding the collective behavior of people in a society, in terms of their opinions, attitudes or decisions, is a very complex but at the same time fascinating goal of sociophysics.  As this research field is becoming more mature, is facing many important and difficult challenges that could inspire novel research. \\

Modern sociophysics has been largely developed by statistical physicists that apply different agent-based models to study various social phenomena such as opinion formation, cultural dissemination, neighborhood segregation, language competition, crowd behavior, political polarization and rumor spreading, among others \cite{Castellano-2009}.  The Ising \cite{Galam-1982}, voter \cite{Holley-1975,Redner-2019}, majority rule \cite{Galam-2002,Krapivsky-2003} and Sznajd \cite{Sznajd-2000,Stauffer-2000} models are among the most studied families of models for discrete opinion dynamics.  They basically consist of a population of agents that can hold one of two possible opinions, for instance to be in favor or against a given issue, represented by up and down spins.  An agent can change its opinion (spin flip) by interacting with its neighbors in a given topology (square lattice, complex network, etc), following a dynamics like Metropolis Monte Carlo, simple neighbor's imitation or majority rule, among others.  These models allow to explore in a oversimplified manner the conditions under which a population of interacting agents can reach a collective state of consensus where all agents share the same opinion (order), or a coexistence of different opinions that describes a fragmented population (disorder).  When agents are endowed with more than two opinion states the population could evolve to a polarized state, with the emergence of two groups of agents that adopt opposite and extreme opinions on a given spectrum \cite{Mas-2013,LaRocca-2014}. 

Continuous opinion models describe a situation where the positions of the individuals vary smoothly within a range of possible choices, like their political orientation, and so their opinions are represented by a real number from $-1$ (extreme left) to $1$ (extreme right).  Each agent can interact only with those agents that are less than a distance apart (threshold) in the opinion space \cite{Deffuant-2000,Hegselmann-Krause-2002,Lorenz-2007}, decreasing their initial opinion difference.  These bounded--confidence models display a final state of consensus when the interaction threshold is large enough, while a fragmentation in different non-interacting opinion groups is observed for small thresholds.  Continuous opinions can also evolve without interaction constraints, according to a weighted matrix that captures the social influence between the neighboring agents in a network \cite{Degroot-1974,Anderson-2019}.  The system eventually reaches consensus if the influence network satisfies some connectedness conditions. \\

Despite the fact that these models attracted a lot of attention and were extensively studied by mathematicians, physicists and engineers, they gave very limited novel insight about opinion dynamics in real life, and did not excite sociologists \cite{Schweitzer-2018}.  In the same line, the sociophysics approach to social phenomena in general suffers several weaknesses already pointed out by active researchers in the field \cite{Schweitzer-2018,Fortunato-2013,Flache-2017,Sirbu-2017}.  Thus, part of the Physics community seem to have agreed on several pitfalls that face modern sociophysics, and how to advance in the direction of becoming a discipline that is respectable by other established fields of science, and is also valued by sociologists.  This step forward might be a sign of maturity of the field. \\

First, a big issue is that there are many theoretical models and a dearth of solid empirical observations and social experiments, as already noticed some years ago and pointed out in different later reviews \cite{Castellano-2009,Sobkowicz-2009,Fortunato-2013,Flache-2017,Sirbu-2017,Peralta-2022}.
This imbalance should be corrected by performing more empirical works.  There is a need for collecting empirical data from social experiments that would allow to test assumptions about social interactions at the microscopic level and to test predictions at the macroscopic level.  Fortunately, this is started to happen.  For instance the authors in \cite{Chacoma-2015} carried out computerized experiments with human subjects to explore the effects of people's influence on opinion formation.  Besides its opinion (answer to a question), each subject states its confidence in it and interacts with other participants in various rounds.  The following types of behaviors were observed, ordered by their frequencies: (i) keeping the original opinion ($60 \%$), (ii) approaching the reference opinion ($30 \%$), and (iii) adopting the reference opinion ($10 \%$).  In other laboratory experiment with people, the authors in \cite{Mas-2013} explored the social influence mechanisms that could lead to opinion bi-polarization.  The results supported the idea that the exchange of arguments during an interaction between participants combined with homophily gives rise to bi-polarization, even in the absence of negative influence.  Furthermore, they showed that social influence based only on the opinion value of the interacting subjects did not affect their opinion difference. 

Second, theoretical models describing a given social phenomena have started to be challenged by new empirical data, mainly coming from the analysis of big data performed by social scientists.  Models require calibration as well as validation to be taken as valid descriptions of real social phenomena. In simple words, ``models cannot avoid validation anymore", as pointed out in \cite{Fortunato-2013}. 

Third, even though reliable model prediction and forecasting of social phenomena is still not possible, theoretical models need to incorporate more realistic features of social interactions, in an attempt to increase their predictive power.  These realistic features should be based either on known mechanisms in social psychology or experimental evidence.  Besides, different models usually implement different rules that lead to the same patterns of a given phenomena, like opinion formation.  These models have to be compared, related and integrated into a single model that describes a particular phenomenon \cite{Flache-2017}.  This can already be done in opinion dynamics because of the large amount of different models and their variants, and it could also be done in other topics of social phenomena. \\

The relative recent availability of the so-called Big data through social networks like Facebook and Twitter is providing data scientists with valuable information about the online behavior of large volumes of internet users.  This also brought new challenges for theoretical physicists that try to understand that data.  Many works in sociophysics are nowadays related to computational social science \cite{Lazer-2009}, a science field that approaches social phenomena from a data-driven perspective, searching for social patterns and universal statistical laws.  For instance, it was found that the time interval between two consecutive messages in online chat rooms follows a power-law and a stretched-exponential distributions \cite{Garas-2012}.  Another example is the study of collective attention in twitter \cite{Lehmann-2012}, which revealed that the evolution of the popularity of hashtags can be classified in four different classes, depending on how the activity behaves around the popularity peak: either activity concentrated before and during the peak, or during and after the peak, or symmetrically around the peak, or on a single day. Also, early works on the study of electoral performance of candidates in an election found that the distribution of votes is well described by a log-normal distribution, which can be explained by means of a multiplicative process \cite{Fortunato-2007}.  The previous examples show how physicists were able to identified statistical regularities at the collective level, and also the dynamic mechanisms that reproduce them by using simple models. \\

Latest Big data trends suggest that computational social science is emerging as a new kind of social science that is entirely based on data analysis \cite{Schweitzer-2018}.  This approach to understanding social phenomena should be supplemented by the theoretical modeling approach that is rooted in traditional science, and searches for fundamental research questions.  That is, computational social science techniques can detect statistical correlations that reveal social patterns by processing large amounts of data, but are less useful to formulate questions.  That is why a physics viewpoint is needed, not only for developing questions, but also for establishing cause-effect relations and for determining emergent consequences at a collective level of a given interaction mechanism at the micro level \cite{SanMiguel-2020}. \\

Developing models that tackle these issues is one of the major technical challenges that sociophysicists must face.  This natural journey have already been taken by other interdisciplinary fields such as biophysics, with successful results, and there is no reason why it cannot be taken by sociophysics. \\

The special issue “Statistical Physics of Opinion formation and Social Phenomena” aims to cover recent advances and present novel theoretical and numerical techniques for the study of social systems, using agent-based models, data analysis, dynamical systems, game theory and Monte Carlo simulations, among others.  We aim to encourage researchers to contribute in the possible research directions described above: empirical works, data analysis, agent-based models with realistic features, calibration and validation against real data.  We would also like to give the opportunity to show the topics that researchers in sociophysics and computational social science are interested in, with the aim of having an overview of the new trends.  We are also interested in collecting works in both theoretical models and data analysis, to make aware researchers of the works that are being done in each field.  Finally, as similar ideas and studies on opinion dynamics are being carried out not only in the physics and mathematics communities but also in the systems and control engineering community \cite{Anderson-2019} using different models and approaches, we want to make their works visible to eventually foster cross-collaborations.

\begin{acknowledgments}
The author acknowledges financial support from CONICET (Grant No. PIP 0443/2014) and from Agencia Nacional de Promoci\'on  Cient\'ifica  y Tecnol\'ogica  (GrantNo. PICT 2016 Nro 201 0215).
\end{acknowledgments}

\bibliographystyle{apsrev}

\bibliography{references}

\end{document}